# Characterization of TiAlSiON Coatings Deposited by Plasma Enhanced Magnetron Sputtering: XRD, XPS, and DFT Studies

A.S. Kamenetskih[a], A.I. Kukharenko[b,c], E.Z. Kurmaev[b], N.A. Skorikov[b], N.V. Gavrilov[a], S.O. Cholakh[c], A.V. Chukin[c], V.M. Zainullina[b,d], and M.A. Korotin[b,*]

[a]*Institute of Electrophysics, Russian Academy of Sciences, Ural Division, Amundsen Street 106, 620016 Yekaterinburg, Russia*
[b]*Institute of Metal Physics, Russian Academy of Sciences, Ural Division, S. Kovalevskoi Street 18, 620990 Yekaterinburg, Russia*
[c]*Ural Federal University, Mira Street 19, 620002 Yekaterinburg, Russia*
[d]*Institute of Solid State Chemistry, Russian Academy of Sciences, Ural Division, Pervomajskaya Street 91, 620990 Yekaterinburg, Russia*
[*]*E-mail: michael.korotin@imp.uran.ru*

**Abstract.**

The results of characterization of TiAlSiON hard coatings deposited on ferric-chromium AISI 430 stainless steel by plasma enhanced magnetron sputtering are presented. The coating with maximum hardness (of 45 GPa) was obtained at the following optimal values of elemental concentrations: Si ~5 at.%, Al ~15 at.%, and Ti ~27 at.%. The elastic modulus of the coating was 590 GPa. The reading of gaseous mixture (Ar-$N_2$) pressure was $1 \cdot 10^{-3}$ Torr and the reading of partial pressure of oxygen ($O_2$) was $1 \cdot 10^{-5}$ Torr. The X-ray diffraction (XRD) measurements showed the presence of Ti(Al)N. High-energy resolved XPS spectra of core levels revealed the formation of Ti-N, Ti-O-N, Si-N and Al-O-N bonds. Comparison of XPS valence band spectra with specially performed density functional theory calculations for two ordered and few disordered $TiN_{1-x}O_x$ ($0 \leq x \leq 1$) demonstrates that a $Ti(Al)O_xN_y$ phase is formed on the surface of AISI430 steel upon the plasma enhanced magnetron sputtering, which provides this material with a good combination of high hardness and improved oxidation resistance.

## 1. Introduction.

Despite a comparatively low oxidation resistance (the oxidation onset is reported at 500° C [1]), titanium nitride (TiN) coating is used widely in cutting tool applications to increase lifetime and to improve performances of cutting tool. It has been found that adding such elements as Al and Si leads to the formation of TiAlSiN composites, which are characterized by high hardness (>40 GPa) and great oxidation resistance (up to 800° C) [2]. To explain the oxidation resistance of TiAlSiN coatings, the several mechanisms have been suggested. A dense $Al_2O_3$ layer has been reported to be formed on the surface, which serves as a barrier against oxygen internal diffusion [3]. It has also been found that *a*-$Si_3N_4$-boundary phase in TiAlSiN coatings prevents diffusion of oxygen along the grain boundaries [4]. To date, a great body of research in TiAlSiN coatings has been focused mostly on the development of various deposition methods and on studying their microstructure, mechanical properties, and oxidation resistance [5-8].

In principle, there exists another way to enhance corrosion resistance of hard coatings, which involves the formation of metal oxynitride phase of Me(N,O) [9]. In this regard, the present paper studies whether it is possible to form the Ti(Al) oxynitride phase by applying a standard technique for preparing hard TiAlSiN coatings, using plasma enhanced magnetron sputtering under an partial pressure of oxygen at $1 \cdot 10^{-5}$ Torr. In the paper, it is shown that the measurements of XPS valence band spectra in combination with the density functional theory (DFT) calculations both for ordered and disordered phases provide a very useful information about the formation of chemical bonds and electronic structure in such multicomponent systems; and the formation of Ti(Al) oxynitride phase is found in TiAlSiN coatings, which can enhance their oxidation resistance.

## 2. Experimental Part and Calculations

*2.1. Coating preparation*

The TiAlSiON films were deposited on AISI 430 stainless steel and T15K6 hard metal by reactive DC magnetron sputtering of targets made of titanium (Ti), aluminum (Al) and silicon (Si) of high purity (Ti - 99,7 %, Al - 99,5 %, Si - 99,999 %) in a $N_2$-Ar-$O_2$ gaseous mixture (Ar - 99,998 %, $N_2$ - 99,999 %). The deposition system was a cylindrical reactor with four flat magnetrons with the targets of 70 mm in diameter. Magnetrons were positioned on the lateral face of the reactor with an internal diameter of 340 mm. Ion assistance with an adjustable density of the ion current was provided by ionizing the gaseous mixture with a wide (50 cm$^2$) electron beam [10]. For this purpose, a low-energy (~100 eV) electron source with a grid plasma cathode [11] was used, which was positioned on the overhead cover of the reactor. Electron emitting plasma in the electron source was generated by a discharge system with a hollow self-heated cathode, which functioned in a combined mode with direct discharge current of up to 7 A and superimposed dc-pulsed current (250 Hz, 200 μs) with an amplitude of up to 100 A. An accelerating voltage of 100 V was applied between a mesh electrode of the electron source (grid) and earthed walls of the reactor. Electrons were accelerated in a double layer of a space charge between the plasma of the electron emitter and the beam plasma. A pulsed (50 kHz, 10 μs, 100 V) negative bias voltage was applied between the samples and the reactor walls. Ion current density from the beam-generated plasma was controlled by change of electron beam current and reaches 60 mA/cm$^2$ in a maximum while average value of ion current density was 4 mA/cm$^2$. The reactor was pumped out using a turbomolecular pump with a pumping speed of 500 l/s. The gaseous mixture was fed to the reactor through the electron source. At a gas flow of 40 ml/min, the total pressure was ~$1\cdot10^{-3}$ Torr. The substrates were disposed on a manipulator with the planetary gyration. The speed of sample rotation relative to the axis of the reactor was 10 rpm. Prior to deposition, the substrates were sputter-etched for 10 min in an Ar atmosphere with 500 eV Ar ions. A Ti adhesion layer of approximately 100 nm in thickness was deposited onto these samples before coating deposition. Elemental compositions of coatings were controlled by changing the discharge currents for magnetrons with Si ($I_{Si}$) and Al ($I_{Al}$) targets and by varying the flow ratio for N$_2$ ($Q_{N2}$) and Ar ($Q_{Ar}$). The partial pressure of O$_2$ was maintained constant (at $1\cdot10^{-5}$ Torr). The discharge current of each of the two magnetrons with Ti targets was constant and equal to 2 A. The coatings weres 1,5-2 μm in thickness. The temperature of samples at all stages of coating deposition did not exceed 300$^0$ C.

*2.2. Coating Characterization*

Hardness measurements of the coatings were performed by nano-indentation using an ultra micro indentation system DUH-211/211S (Shimadzu). The hardness value was obtained by Oliver-Pharr method [12]. Chemical composition of coatings was estimated by EDX (Energy Dispersive X-ray analysis) using a nitrogen-free Aztec X-MAX80 system (Oxford Instruments). For structural analysis, the X-ray diffraction (XRD) measurements were carried out using a Shimadzu XRD 7000 diffractometer using Cu K$\alpha$ radiation and a graphite monochromator. XRD patterns were processed by the method of integral analysis using XPert High Score Plus software. The average size of the coherent scattering regions was estimated by Scherrer method.

XPS core-level and valence-band spectra measurements were made using PHI XPS Versaprobe 5000 spectrometer (ULVAC-Physical Electronics, USA). The Versaprobe 5000 is based on the classic X-ray optic scheme with a hemispherical quartz monochromator and an energy analyzer working in the range of binding energies from 0 to 1500 eV. The greatest advantage of this XPS system is electrostatic focusing and magnetic screening. As a result, the achieved energy resolution is $\Delta E \leq 0.5$ eV for Al *K$\alpha$* excitation (1486.6 eV). The vacuum circuit of Analytical Chamber uses oil-free rotary pump which allow to obtain and keep the pressure not less than $10^{-7}$ Pa. The dual-channel neutralizer (ULVAC-PHI patent) was applied in order to compensate the local charging of the sample under study due to the loss of photoelectrons during XPS measurements. All samples under study were previously kept in the Intro Chamber within 24 hours under rotary pumping. After that the sample was introduced into the Analytical Chamber and controlled with the help of "Chemical State Mapping" mode in order to detect micro impurities. If the micro impurities were detected then the sample was replaced from the reserved batch. The XPS spectra of core-levels and valence band were recorded in Al *K$\alpha$* 100 μm spot mode with x-ray power load of the sample less than 25 Watts. Typical signal-to-noise ratio values in this case were not less than 10000/3. The spectra were processed using ULVAC-PHI MultiPak Software 9.3.

*2.3. Electronic structure calculations*

The experimental lattice constant $a$=4.2433 Å for the cubic cell of TiN (space group *Fm-3m*, Z=1) was taken from [13]. The pseudo-potential Quantum-Espresso code [14] was used to calculate the band structure of titanium oxynitride in ordered phases. We used the ultrasoft pseudo-potential with Perdew-Zunger (PZ) version of exchange potential [15]. The plane-wave and kinetic energy cut-off's were chosen to be 50 Ry and 200 Ry, respectively. Integration over the Brillouin zone was performed over 16x16x16 k-point mesh.

Two approaches were used to model doping of TiN by oxygen. First one was a supercell approximation, second one − the coherent potential approximation. For the first approach the constructed supercell contained four formula units of TiN, where one or two of nitrogen atoms were replaced by oxygen. So calculations were performed for $TiN_3O_1$ and $TiN_2O_2$ compounds. In the second approach arbitrary concentration of oxygen impurity in the nitrogen site is calculated for the unit cell of TiN.

The basic equations of the coherent potential approximation used in the present investigation are described in [16]. For self-consistent calculation of the coherent potential, it is necessary to build a Hamiltonian $H_0$ of the undoped system and to determine parameters $\Delta V$ describing the difference between the impurity and the host sites. The Hamiltonian $H_0$ was calculated in frames of the linearized muffin-tin orbital method in the tight binding approximation (TB LMTO) [17] and projected into Wannier functions basis according to the procedure presented in [18]. This Wannier functions basis included *4s*- and *3d*-orbitals of Ti atom together with *2s*- and *2p*-orbitals of N atom. The $\Delta V$ parameters for *s*- and *p*-functions were calculated to be −6.47 eV and −3.33 eV, correspondingly. The CPA calculations were performed with the inverse temperature value β=10 eV, Matsubara frequencies were cut off at 1000 eV.

# 3. Results and Discussion

## 3.1. Mechanical properties

Figure 1 displays hardness of the coatings as a function of $I_{Si}$ (at constant values of $I_{Al}$=2.6 A and $Q_{N2}$=12 sccm, $I_{Al}$ ($I_{Si}$=0.7 A, $Q_{N2}$=12 sccm) and $Q_{N2}$ ($I_{Si}$=0.75 A, $I_{Al}$=2.5 A).

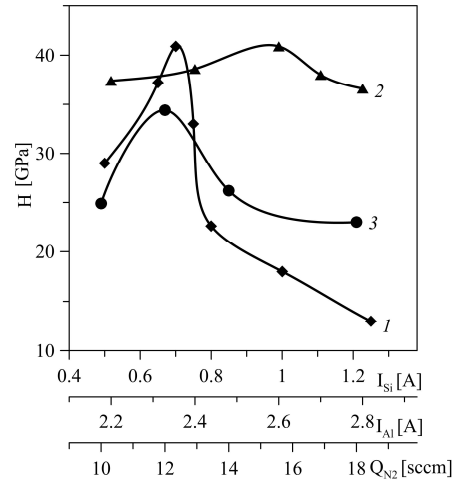

Fig. 1. The nanohardness of coatings as a function of discharge current of magnetrons with Si (*1*), Al (*2*) targets and N$_2$ flow rate. The pulse ion current density was 60 mA/cm$^2$.

As shown in Fig.1, Si content, which increases from 3 to ~10 at. % with $I_{Si}$ increasing from 0.5 A to 1.25 A, causes the most essential influence on hardness of the coatings. The hardness reaches its maximum value of 41 GPa at Si content of nearly 5 at.% and decreases sharply as Si content increases further. Changing Al content in the range of 11-19 at.% at $I_{Al}$ increasing from 2.2 A to 2.8 A is accompanied with a change in the coatings hardness by no more than 4 GPa. The coating with the maximum hardness (41 GPa) was obtained at the following optimal values of elemental contents: Si~5 at.%, Al~15 at.%, Ti~27 at.%; and at the value of $Q_{N2}$=12 sccm. The elastic modulus of the coating was 590 GPa.

Annealing of the coatings in vacuum at 800° C during 2 h leads to an increase in hardness by 2-7 GPa. The most essential growth of hardness corresponds to Si contents below the optimal value (~3 at. %). The maximum hardness (43 GPa) was achieved upon annealing.

Coatings deposited at the optimal parameters ($I_{Si}$=0.7 A, $I_{Al}$=2.6 A, $Q_{N2}$=12 sccm) and different amplitudes of electron beam current were investigated in order to evaluate the influence of pulsed ion current density on the coating characteristics. The mean current of electron beam was kept constant (at 7 A). To keep the value of mean electron beam current constant at decreasing pulse current, the direct current was raised. Hardness of the coatings as a function of electron beam current amplitude is shown in Fig. 2. It was established that as hardness of the coatings increased by 10 GPa with the pulse ion current density increasing from 0 to 60 mA/cm$^2$ (increasing the amplitude of electron beam current up to 100 A).

An increase in the coatings hardness from 31 to 45 GPa has been observed earlier at the ion/atom arrival ratio fluxes changing from 2.5 to 4 in the DC mode [19]. Enhancement of mechanical properties can be explained by an increase in the species mobility in the growing film thus leading to a better phase segregation. Unlike [19], in the present work the coatings were deposited at a constant value of the mean ion current density, and the increasing pulse ion current density appears to be the most probable reason for the hardness growth.

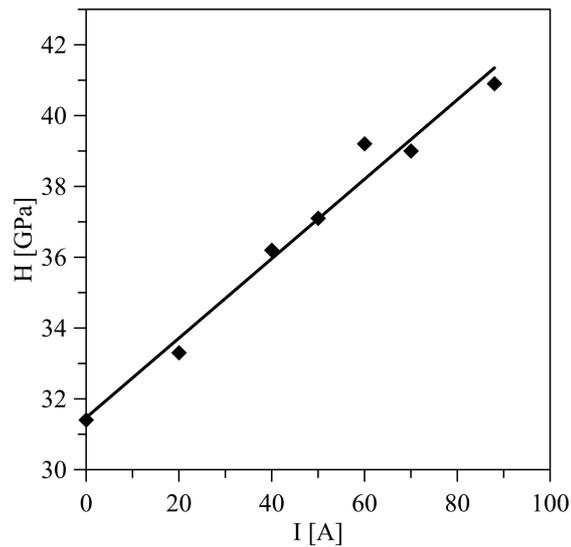

Fig. 2. The nanohardness of coatings as a function of amplitudes of electron beam current.

3.2. XRD analysis

XRD patterns of samples coated at different values of pulsed electron beam current are shown in Fig. 3. The main phase is the solid solution of Ti(Al)N with a fcc lattice of NaCl-type. Reflexes of α-Fe (2θ~44.7°) represented on the XRD pattern correspond to the substrate material. Characteristic peaks of the main phase of Ti(Al)N at 2$\theta$~36.7°; 42.6°; 61.8° correspond to the crystalline planes (111), (200) and (220). The preferred orientation of grains is (200).

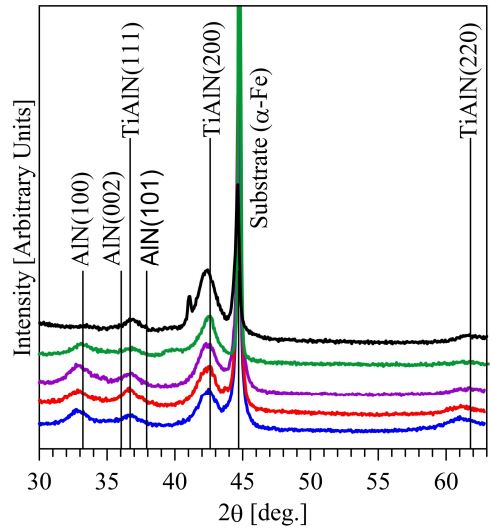

Fig. 3: XRD patterns of AISI430 samples coated with TiAlSiON at amplitudes of the electron beam current: *1*–20, *2*–40, *3*–50, *4*–70, and *5*–100 A.

The peak with the maximum at 2$\theta$~33.2° can be attributed to AlN phase with (100) texture [20]. Other peaks of AlN phase at 2$\theta$~36.0° and 37.9° are not observed because of a low intensity or a strong texture of the phase. The AlN peak intensity decreases as the amplitude of electron beam current increases up to 100 A, which is indicative of the reduction in the AlN phase fraction in the coating. Thus, a conclusion can be made that the increased hardness of the coatings with the increase in the pulse ion current density can be as connected with the enhanced phase segregation as caused by the solid-solution hardening [21].

*3.3. XPS spectra*

The XPS survey spectra (Fig. 4) show that the surface of as-prepared sample is strongly contaminated with oxygen and carbon. After Ar$^+$ ion etching (V=2 keV, $\tau$=2 min) the surface cleaning takes place, the oxygen signal is strongly reduced, the carbon signal is almost disappeared and mainly signals from Ti, N, Al and Si from TiAlSiON coatings are recorded. The surface composition estimated from the XPS survey spectra is given in Table 1.

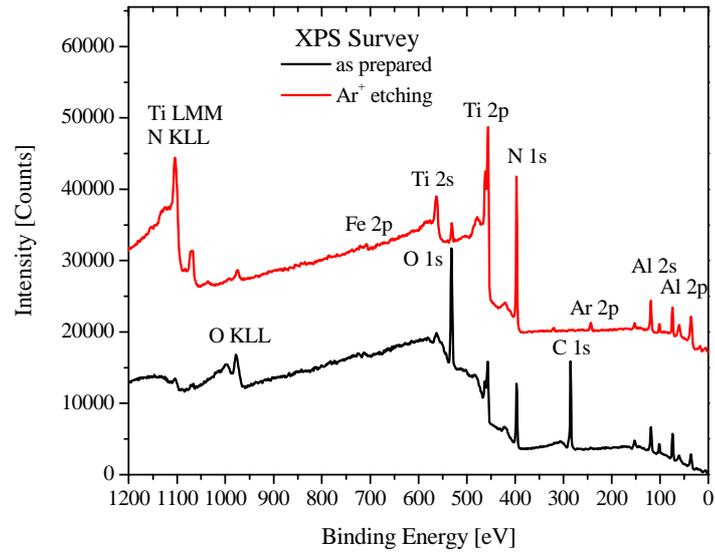

Fig. 4. XPS survey spectra of TiAlSiON coatings.

Table 1. Surface composition in at. %.

| XPS line | C $1s$ | O $1s$ | N $1s$ | Al $2p$ | Ti $2p$ | Si $2p$ | Fe $2p$ | Ar $3p$ |
|---|---|---|---|---|---|---|---|---|
| as prepared | 37.2 | 26.0 | 17.6 | 10.5 | 6.1 | 2.4 | 0.1 | - |
| after Ar$^+$ etching | 0.3 | 4.5 | 48.9 | 16.2 | 24,5 | 2.8 | 1.3 | 1.4 |

The high-energy resolved XPS spectra of core-levels (Figs. 5-6) show that Fe in an as-prepared sample (from AISI430 substrate) is strongly oxidized and has the $Fe^{3+}$ oxidation state. After $Ar^+$ ion etching, Fe is reduced up to the metallic state. The Ti $2p$-spectra of as-prepared sample have a three-peak structure at 455.0, 456.6 and 458.3 eV, which can be attributed to Ti-N, Ti-O-N and Ti-O bonds, respectively [22, 23]. The Ti-O contribution corresponds to the formation of $TiO_2$ oxide layer on the surface and is not observed after $Ar^+$-ion etching.

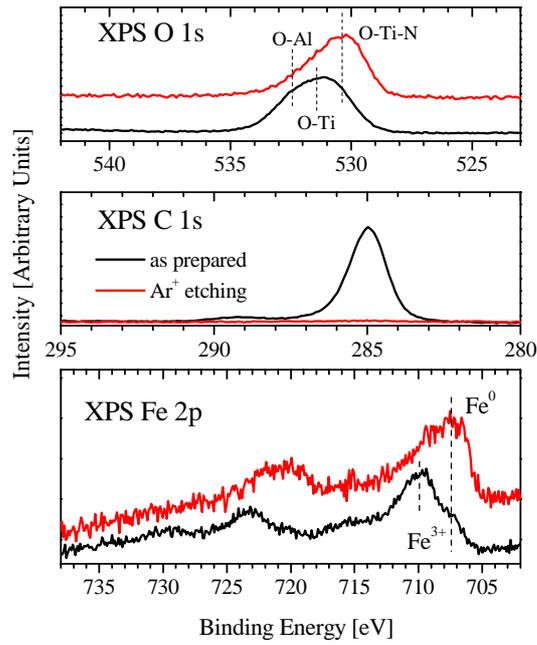

Fig. 5. XPS C *1s*, O *1s* and Fe *2p* spectra of TiAlSiON coatings.

XPS Al *2p* spectrum of as-prepared TiAlSiON coating consists of contributions of Al-Al, Al-N and Al-O bonds (Fig. 6). After $Ar^+$-ion etching the Al *2p*-spectrum becomes narrower with the prevalence of Al-N and Al-O bonds. The binding energy of XPS N *1s* spectra for both as-prepared and $Ar^+$-ion etched samples show the presence of both Ti-N and Al-N bonds.

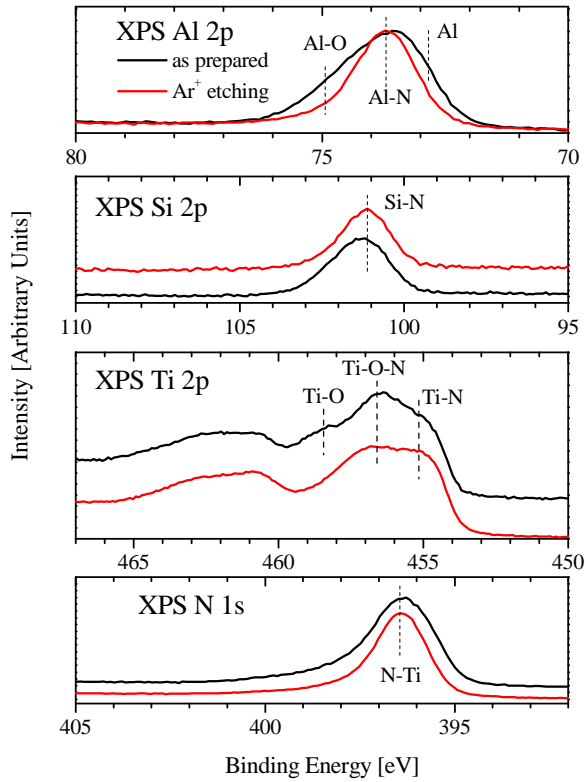

Fig. 6. XPS Al *2p*, Si *2p*, Ti *2p* and N *1s*- spectra of TiAlSiON coating.

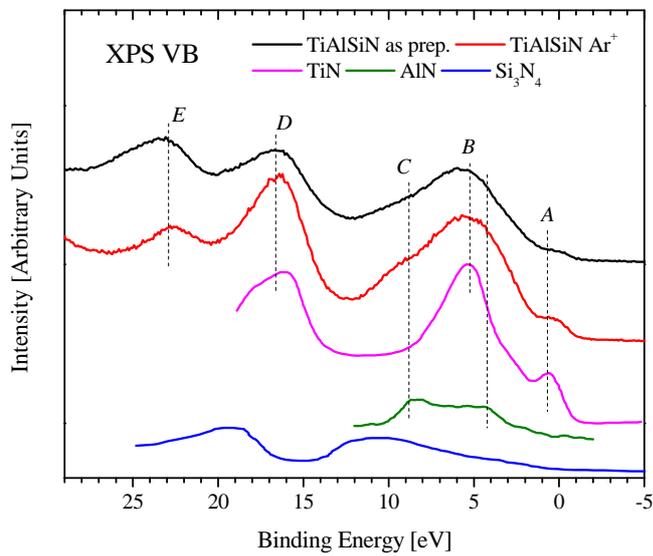

Fig. 7. The comparison of XPS spectra of TiAlSiON coating with spectra of reference samples: TiN [24], $Si_3N_4$ [25] and AlN [26].

XPS valence band spectra of TiAlSiN coatings are presented in Fig. 7 together with the spectra of reference samples TiN, $Si_3N_4$ and AlN. One can see the resemblance of the *B* and *C* features of TiN and AlN spectra in energy with those of TiAlSiN, whereas the spectrum of $Si_3N_4$ is found to be different. For analogy with XPS VBs of TiN and $TiO_2$, the low energy peaks at ~22.9 (*E*) and 16.6 eV (*D*) of TiAlSiN coatings can be attributed to the formation of O *2s* and N *2s*-subbands, respectively. The comparison of XPS VBs of TiAlSiON and TiN show that the features of *D*, *B* and *A* are close in energy for both compounds, which means that these features are mainly due to the contributions of N *2s*, N *2p* and Ti *3d*-states. It is interesting to note that the energy difference between *D* and *B* peaks (~10.9 eV) is close to the difference of *2s* and *2p* orbital energies for a nitrogen atom [27], which confirms our interpretation of the origin of *D* and *B* peaks in XPS VBs of TiAlSiN coatings. Driven by the same logics, we can estimate the energy position of O *2p*-derived band. According to Ref. 27, the calculated energy difference of O *2s* and O *2p* orbital energies is 15.0 eV, which is close to the energy difference between *E* and *C* peaks in XPS VB of TiAlSiON, which allows the origin of these peaks to be attributed to the contribution of O *2s* and O *2p*-states, respectively. Therefore, our analysis of XPS VB spectra of TiAlSiON coatings and their comparison with the spectra of reference samples allows us to suggest the formation of Ti-Al-O-N phase with NaCl structure with partial substitution of Ti and N atoms by Al and O, respectively.

*3.4. DFT calculations*

The results of calculations within the Density Functional Theory (DFT) of the electronic structure of $Ti(N_{0.5}O_{0.5})$ and $Ti(N_{3/4}O_{1/4})$ supercells are presented in Fig. 8. According to these DFT calculations, the low energy subbands located at ~20.9 (*E*) and 15.2 eV (*D*) are formed by O *2s* and N *2s*-states, respectively. The main valence band consists of three subbands (*C*, *B* and *A*) centered at ~7.6, 5.5 and 1.2 eV, respectively, which arise from the main contributions of O *2p*, N *2p* and Ti *3d*-states. As seen, these data are in a good agreement with XPS valence band measurements (Fig. 7). Looking for the relative ratio of *B* and *C* peaks in experimental valence band spectra and calculated electronic structure of $TiO_{1/2}N_{1/2}$ and $TiO_{3/4}N_{1/4}$ one can conclude that ~¼ of oxygen atoms is substituted by nitrogen atoms and form the Ti-Al oxynitride phase.

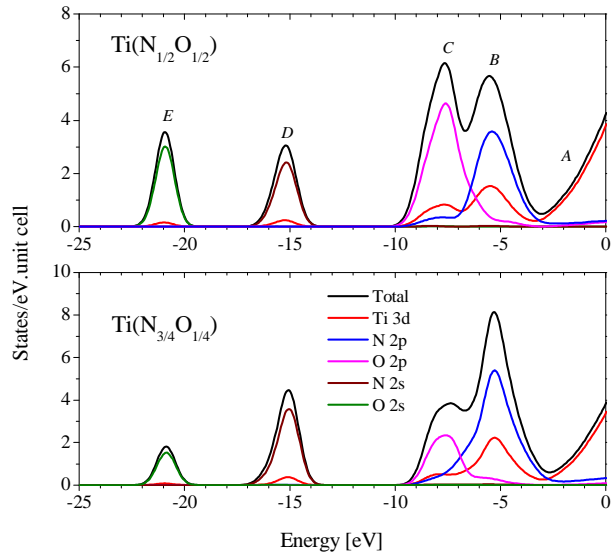

Fig. 8. Calculated total and partial density of states for $Ti(N_{1/2}O_{1/2})$ and $Ti(N_{3/4}O_{1/4})$.

Further detailing of concentration dependence of electronic spectra of disordered titanium oxynitrides $TiN_{1-x}O_x$ was carrying out within the coherent potential approximation for wide range of $x$=0.1, 0.2, …, 0.9, 1.0. The electronic spectrum alteration obtained in these CPA calculations is presented in Fig. 9.

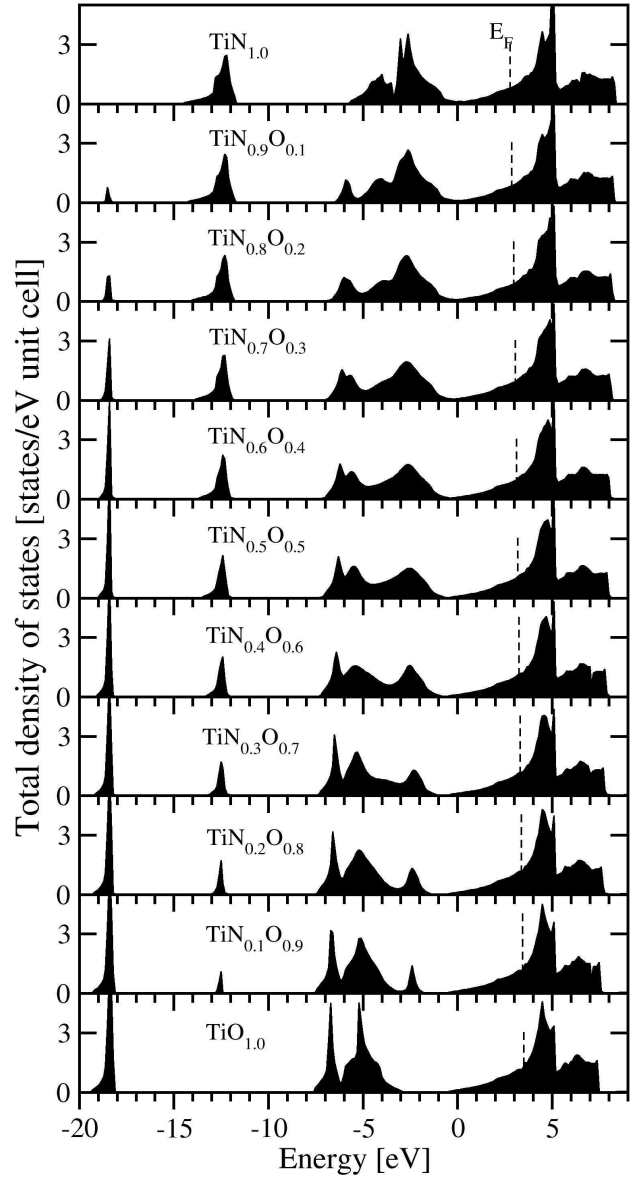

Fig. 9. Total densities of states for disordered $TiN_{1-x}O_x$ calculated within the coherent potential approximation. Fermi level is shown by vertical dashed line.

The electronic structures of disordered $TiN_{1-x}O_x$ have the same 5-sub-bands composition as it was determined in pseudopotential calculations. With the increase of $x$, the appearance, intensity increase and broadening of the oxygen $2s$-band is observed. At the same time the intensity of the nitrogen $2s$-band decreases. Thus, the ratio of O$2s$/N$2s$-bands intensities could be a tool for the evaluation of the impurity content in titanium oxynitrid samples. The best correspondence between experimental and theoretical intensity ratios of corresponding bands is found for the composition $TiN_{0.85}O_{0.15}$ (Fig. 10).

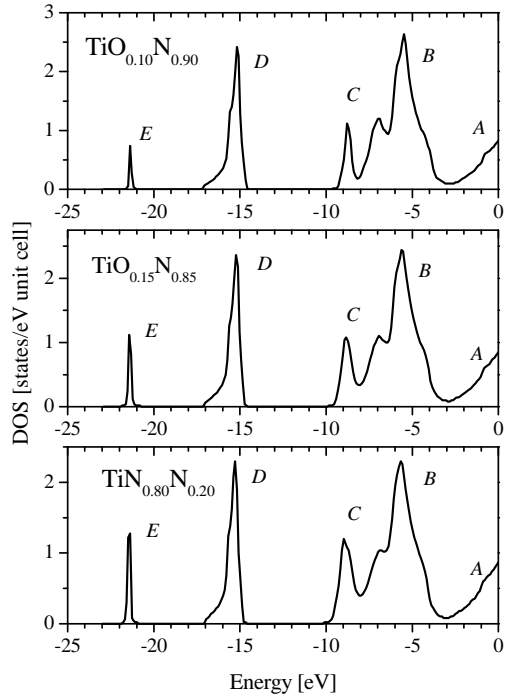

Fig. 10. Total densities of states for disordered TiO$_x$N$_{1-x}$ with $x$=0.10, 015 and 0.20.

It seems that this result allows us to identify the formation of Ti(Al)ON phase in TiAlSiON coatings. In addition to Al$_2$O$_3$ barrier against oxygen internal diffusion and $a$-Si$_3$N$_4$-boundary phase, which prevents the diffusion of oxygen along the grain boundaries, the formation of Ti-oxynitride phase is also favorable for corrosion resistance of TiAlSiN coatings (see Ref. 9).

## 4. Summary

The TiAlSiON coatings were deposited on AISI430 stainless steel by plasma enhanced magnetron sputtering with hardness of 41 GPa and elastic modulus of 590 GPa. XRD measurements have shown the presence of Ti(Al)N-based phase with NaCl structure and AlN phase. The high-energy resolved XPS measurements of core levels (Ti $2p$, Al $2p$, N $1s$ and O $1s$) revealed the formation of Ti-N, Ti-N-O and Al-O-N bonds. Based on the comparison of XPS valence band spectra with electronic structure calculations of ordered and disordered titanium oxynitrides $TiN_{1-x}O_x$ it is concluded that (Ti-Al)ON phase is formed, which can provide additional oxidation resistance of TiAlSiON coatings.

**Acknowledgements**

The results of experimental measurements of XPS spectra and calculations of electronic structure were supported by the grant of the Russian Scientific Foundation (project no. 14-22-00004).